\documentclass{article}
\usepackage{spconf,amsmath,graphicx,hyperref}
\usepackage{subcaption}
\usepackage{multirow}
\usepackage{threeparttable}

\title{Dynamically Slimmable Speech Enhancement Network with Metric-Guided Training}
\name{Haixin Zhao, Kaixuan Yang, and Nilesh Madhu}
\address{IDLab, Ghent University - imec, Ghent, Belgium\\
\{haixin.zhao, kaixuan.yang, nilesh.madhu\}@ugent.be
}

\begin{document}
\ninept
\maketitle
\begin{abstract}
To further reduce the complexity of lightweight speech enhancement models, we introduce a gating–based Dynamically Slimmable Network (DSN). The DSN comprises static and dynamic components. For architecture-independent applicability, we introduce distinct dynamic structures targeting the commonly used components, namely, grouped recurrent neural network units, multi-head attention, convolutional, and fully connected layers. A policy module adaptively governs the use of dynamic parts at a frame-wise resolution according to the input signal quality, controlling computational load. We further propose Metric‑Guided Training (MGT) to explicitly guide the policy module in assessing input speech quality. Experimental results demonstrate that the DSN achieves comparable enhancement performance in instrumental metrics to the state-of-the-art lightweight baseline, while using only 73\% of its computational load on average. Evaluations of dynamic component usage ratios indicate that the MGT-DSN can appropriately allocate network resources according to the severity of input signal distortion.

\end{abstract}
\vspace{-0.2cm}
\section{Introduction}
\label{sec:intro}
\vspace{-0.2cm}
In single-channel speech enhancement, deep neural networks have achieved the state-of-the-art (SOTA) performance \cite{das2021fundamentals, lu23e_interspeech}.
However, their increased computational load restricts deployments on resource-constrained edge devices.
To address this, a range of optimisation techniques for lightweight networks, such as pruning \cite{9437977}, quantisation \cite{9287739}, depth-wise separable convolutions \cite{rong2024gtcrn}, perceptually inspired dimensionality reductions \cite{schroeter2022deepfilternet2}, and structural grouping strategies \cite{zhao2025studylightweighttransformerarchitectures, dang2023first} have been investigated.

\textbf{Prior work:} 
Most existing methods rely on a static, one-size-fits-all architecture that disregards the varying speech quality in the input signal.
As a result, allocating the same computational resources may lead to over-provisioning for easier inputs while being insufficient for more challenging cases, yielding suboptimal resource utilisation and ultimately compromising enhancement efficacy.
To address this challenge, various dynamic methods have been proposed to reduce computational load by employing techniques such as skipping \cite{9829920}, early exiting \cite{9413933}, and dynamic network path routing \cite{6797059, you21_interspeech}. In \cite{9413933, 10096897}, early exiting frameworks and gating mechanisms were utilised to adaptively reduce model complexity. However, these approaches typically operate at the sample level and lack finer granularity to account for varying speech quality within samples.
In response, a series of gating-based dynamically slimmable networks have been developed, which adaptively adjust the utilisation of convolutional and fully connected layers based on resource constraints or input characteristics \cite{10248143, 10638203}. 
Skip Recurrent Neural Networks (RNNs) have been proposed to dynamically skip state updates, thereby reducing unnecessary computations \cite{fedorov20_interspeech}. This strategy has been further extended to parallel RNNs to address issues arising from interrupted mask updates \cite{9829920}. However, these frameworks are often limited to a specific network component type.

In most existing dynamic architectures, a policy sub-network derives gating vectors solely from input features and does not leverage explicit information about input quality. Instead, its learning is guided implicitly by the trade‑off between reconstruction quality and resource utilisation \cite{10248143, Veit_2018_ECCV}. 
Introducing voice activity detection is a useful preliminary attempt at exploiting explicit characteristics \cite{9829920}. 
However, speech quality information, which can truly reflect the enhancement difficulty of input, has not been well utilised.

\begin{figure*}
\centerline{\includegraphics[width=1\linewidth]{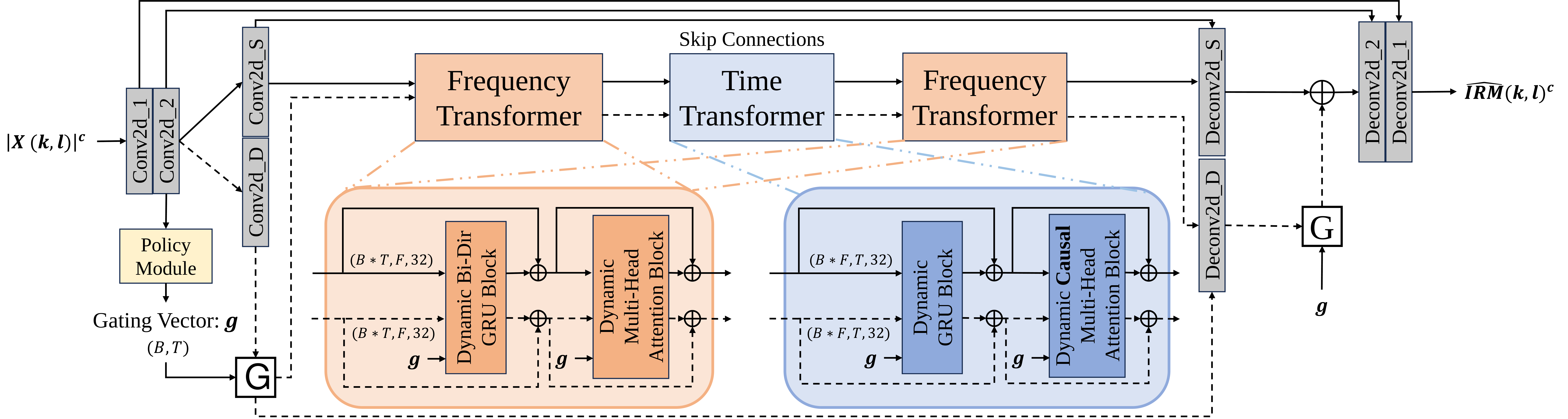}}
 \vspace{-0.2cm}
\caption{The architecture of the Dynamic Slimmable Network, in which the third convolutional layer pair, GRU and MHA modules, are dynamic components. The static network paths, denoted by solid arrows, are always processed, while dynamic paths (dotted lines) can be dynamically slimmed during inference, guided by the estimated frame-wise gating vector $\textbf{g}$. $\textbf{g}$ is identically used for all dynamic modules. B denotes batch size, and T, F, and 32 are the tensor sizes along time, frequency and channel, respectively. Boxed $G$ denotes gating operations. 
}  
\vspace{-0.35cm}
\label{fig1}
\end{figure*}

\begin{figure}
\centerline{\includegraphics[width=1\linewidth]{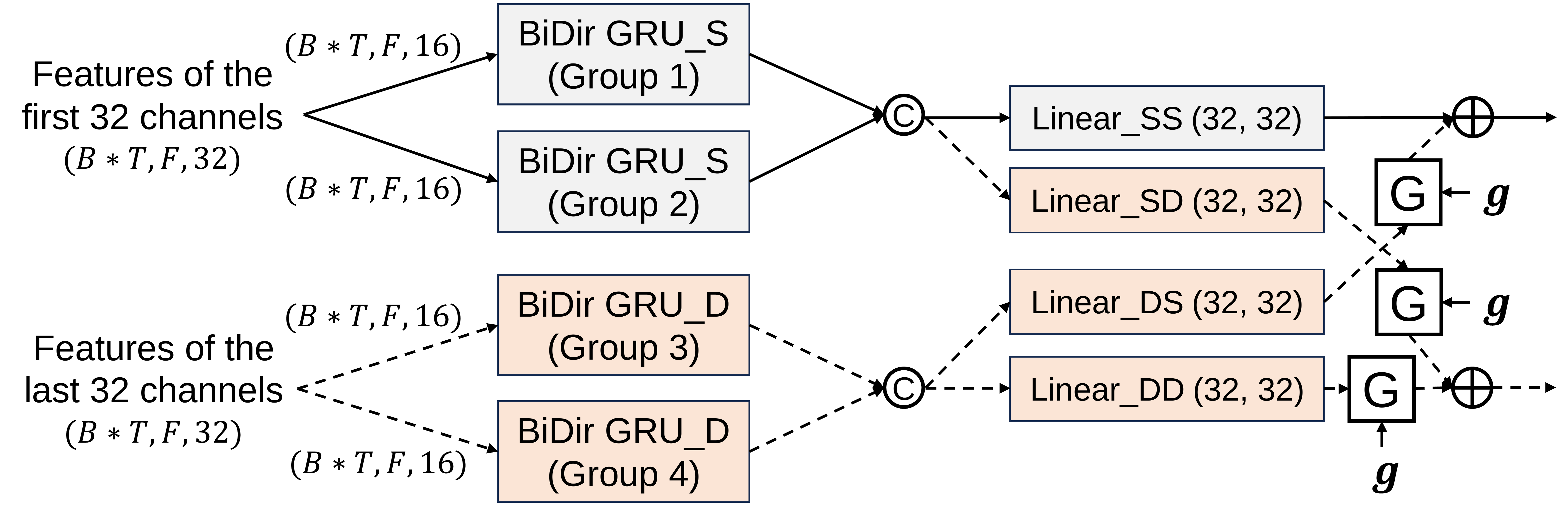}}
 \vspace{-0.25cm}
\caption{The structure of the dynamic GRU block in frequency transformers. Circled $C$ denotes concatenation.}  
\vspace{-0.2cm}
\label{fig3}
\end{figure}

\textbf{Contributions:} 
To address challenges in dynamic frameworks limited to specific network components, we extend the dynamic frameworks applicable to the most commonly used layers in lightweight architectures, including multi-head attention (MHA), grouped RNNs, convolutional, and fully connected layers. A gating-based Dynamically Slimmable Network (DSN) is proposed, which dynamically activates selective network components in a frame-wise manner to further reduce the computational complexity. We benchmark the proposed DSN on the SOTA lightweight model to demonstrate its efficacy.
To further explicitly leverage input characteristics, we introduce Metric-Guided Training (MGT) to the proposed DSN, guiding the policy block in assessing speech quality.
Moreover, we evaluate DSN along with activation ratios, defined as the mean of the gating vector for each sample, across two speech quality measures. These evaluations demonstrate the effectiveness of DSN and the contribution of MGT to performance improvement by aligning dynamic activation ratios with the quality of input speech.

\vspace{-0.2cm}
\section{Methods}
\vspace{-0.2cm}
\subsection{Dynamic Slimmable Architecture}
\vspace{-0.2cm}
Speech enhancement methods aim to reconstruct clean speech $s(n)$ from the noisy input $x(n)$. The proposed dynamic slimmable framework is validated on a causal Frequency-Time-Frequency Network (FTF-Net) \cite{zhao2025studylightweighttransformerarchitectures},
which was recently demonstrated to achieve SOTA performance among lightweight enhancement models.
The FTF-Net adopts widely used network components, such as convolutional layers, grouped Gated Recurrent Units (GRUs), MHA, and fully connected layers, making it a well-suited baseline network to validate the broad applicability of the proposed dynamic slimmable framework on diverse network modules.
Similar to FTF-Net, the DSN operates on the compressed amplitude representation of input signals in the Short-Time Fourier Transform (STFT) domain.
Consistent with prior work \cite{9522648}, 
the compression factor, $c=0.3$, is configured to enhance the estimation of low-energy speech components.
The network outputs the Ideal Ratio Mask (IRM) for the amplitude enhancement in the compressed domain, while retaining the input phase for reconstruction. As shown in \cite{zhao2025studylightweighttransformerarchitectures}, additional phase estimation based on FTF-Net does not yield further improvement.

\begin{figure}
\centerline{\includegraphics[width=0.8\linewidth]{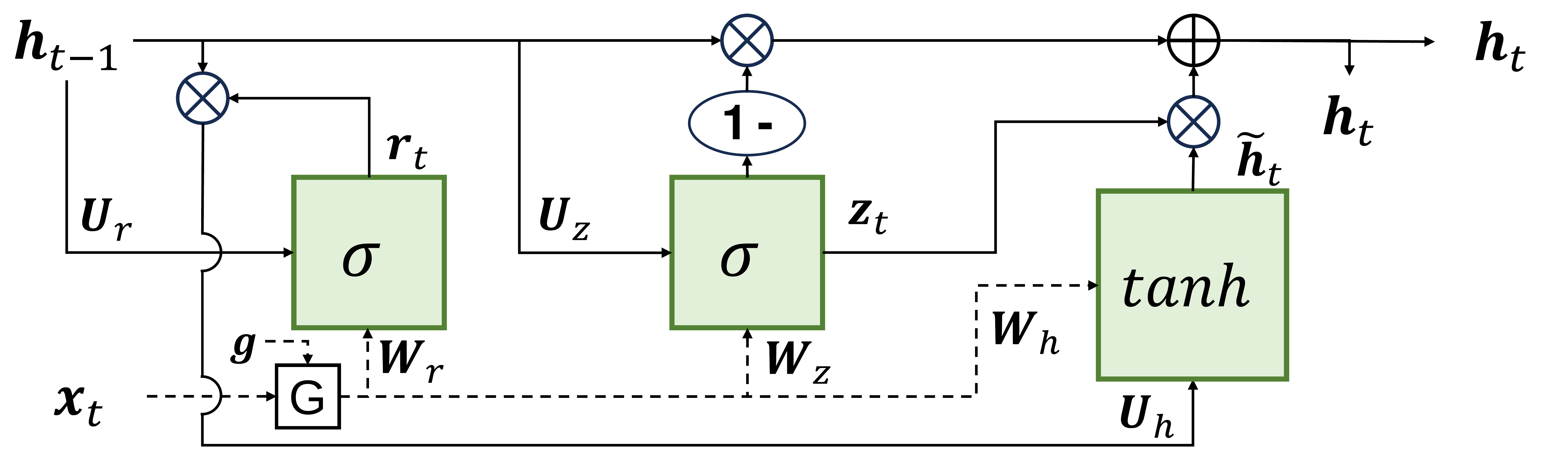}}
 \vspace{-0.2cm}
\caption{The structure of the GRU cell in the time transformer’s dynamic GRU groups. $\boldsymbol{h}$ denotes hidden states, and $\boldsymbol{x}_t$ is input.}  
\vspace{-0.6cm}
\label{fig4}
\end{figure}

As illustrated in Fig.~\ref{fig1}, the proposed Dynamic Slimmable Network preserves the overall 3-layer U-Net structure of the base model. The network begins with two convolutional layers with output channels of 16 and 32, respectively.
The third convolutional layer pair, the GRUs, and the MHA modules are replaced with their corresponding dynamic slimmable counterparts.
A policy module is introduced after the first two convolutional layers to generate a gating vector $\textbf{g}$, where the gating value for frame $t$ is denoted as $g_t$. This vector \textit{globally} guides all gating operations (denoted by the boxed $G$) in dynamic blocks, enabling the dynamically slimmable mechanism.
The policy module first extracts the channel‐wise mean $\boldsymbol{\mu}$ and standard deviation $\boldsymbol{\sigma}$ of features. These statistics are then passed through two stacked fully connected layers with 16 and 2 output channels, respectively, to estimate the logits of a binary distribution per frame. Finally, a Gumbel-Softmax \cite{jang2017categoricalreparameterizationgumbelsoftmax} with a scalar temperature of $\tau=0.5$ is applied to estimate $\textbf{g}$. 
It is configured in soft mode with a value range of $[0,1]$ during training to enable differentiable learning. The mode is switched to hard during inference to obtain binary gating vectors, thereby facilitating the gating mechanism to function as network slimming. The lower temperature $\tau$ provides harder probabilities and improves the sparsity to mitigate the distribution gap between training and inference.

\begin{figure}
\centerline{\includegraphics[width=1\linewidth]{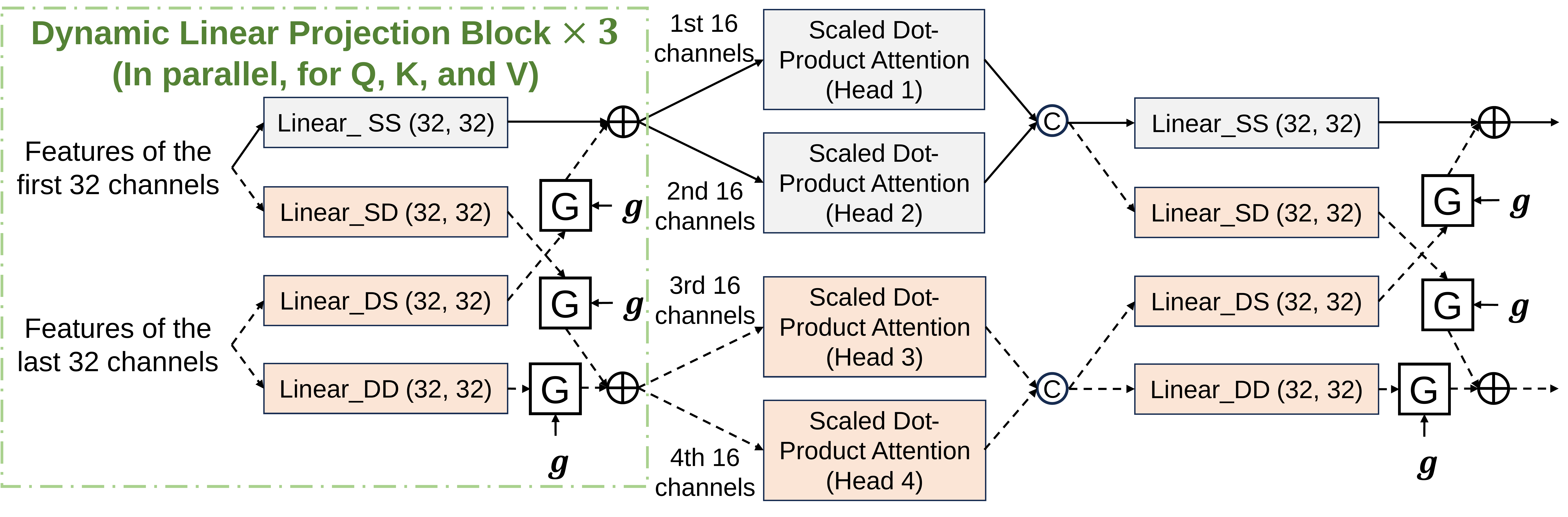}}
 \vspace{-0.2cm}
\caption{The dynamic MHA structure. The parallel linear projections of query, key, and value feature maps share the same dynamic linear block structure but differ in their independent learnable parameters.} 
\vspace{-0.55cm}
\label{fig5}
\end{figure}
\textbf{Dynamic convolutional blocks:} For the third convolutional stage, the feature maps are processed using two parallel convolutional layers, each with 32 output channels. The outputs of these two layers are then routed to the static path and the dynamic slimmable path, respectively. The static $Conv2d\_S$ is always operated, while the dynamic $Conv2d\_D$ supports frame-wise dynamic slimming during inference.
Similarly, two parallel deconvolutional layers are employed in the decoder. As shown in Fig.~\ref{fig1}, the output of the dynamic deconvolutional layer is added to that of the static deconvolutional layer through gating. During inference, the gating vector is discretised to either 0 or 1 for each frame, thereby enforcing a hard selection between using only the static path or combining information from both the static and dynamic paths. 

\begin{figure*}
\centerline{\includegraphics[width=1\linewidth]{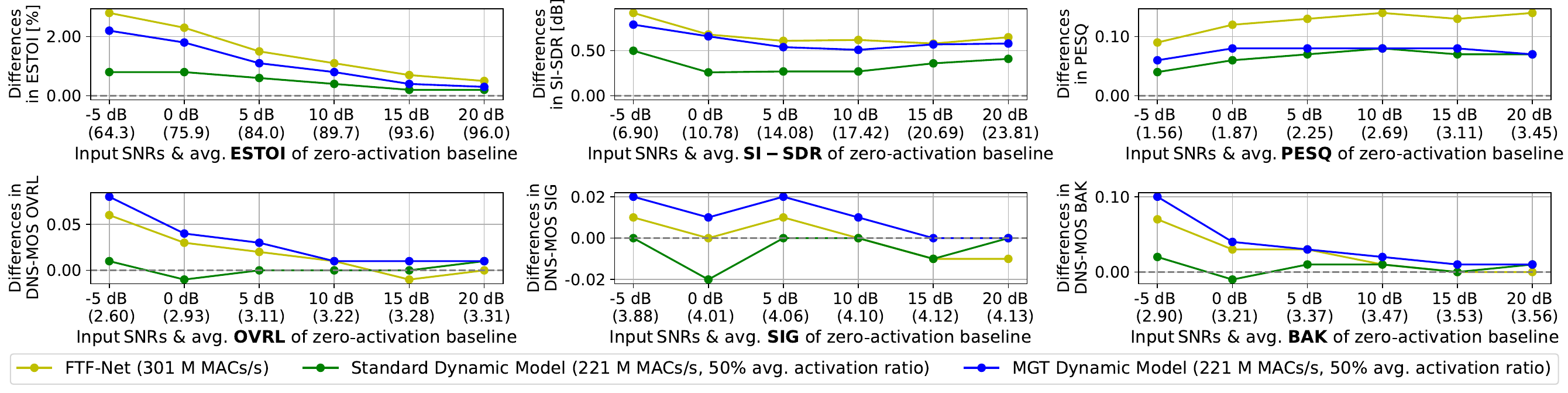}}
 \vspace{-0.25cm}
\caption{
Evaluation results of the proposed dynamic models benchmarked against two static baselines. Performance gains are illustrated relative to the zero-activation baseline, that is, the dynamic model with all dynamic components deactivated, resulting in a computational complexity of 141 M MACs/s. The six-panel plot contrasts the static FFT-Net (equivalent to the 100\% activated dynamic model), the standard dynamic model, and the MGT dynamic model across six instrumental metrics, at SNRs from 5 dB to 20 dB. Both dynamic models exhibit a 50\% average activation ratio on the test dataset, corresponding to a computational complexity of 221 M MACs/s. The average metric score of the zero-activation baseline is indicated below the corresponding SNRs. The average MACs/s of models are provided as well for evaluation.}  
\vspace{-0.52cm}
\label{fig6}
\end{figure*}

\textbf{Dynamic GRU blocks in frequency transformers:}
As illustrated in Fig.~\ref{fig3}, the four bi-directional groups in the frequency transformers are divided into two static and two dynamic ones.
The subsequent fully connected layer, responsible for information exchange across groups, is redesigned by a dynamic linear block with 4 sublayers. Only the information from the static groups is consistently utilised and forwarded to the next static blocks. The information from dynamic groups is weighted and integrated into the static path by the dynamic gating. 
The dynamic path here is formed by combining outputs from \textit{both} static and dynamic groups via two gating operations, as shown in Fig.~\ref{fig3}. This is then further propagated to the subsequent dynamic blocks.

\textbf{Dynamic GRU blocks in the time transformer:}
Note that the above dynamic structure is only applicable to GRUs in the frequency transformer, as they process each frame without relying on sequential information from other frames. In contrast, uni-directional GRUs in the time transformer maintain the temporal dependencies, and the hidden state from the current frame is required by the GRU cells in subsequent frames, even when the dynamic GRU groups of the current frame are deactivated.
To address this, we redesign the GRU cells in the time transformer to be partially dynamic. As shown in Fig.~\ref{fig4}, all input-to-hidden paths (dotted lines) within the current frame are gated and can be slimmed during inference when the gating vector is 0. However, the hidden state is always updated and stored for the next frame to maintain temporal continuity. The structure of the static and dynamic GRU groups, as well as the sub-linear layers, remains consistent with that of the frequency transformer illustrated in Fig.~\ref{fig3}, differing only in that the input feature map is permuted to the shape $(B\times F, T, Channels)$. 

\textbf{Dynamic MHA blocks:}
The dynamic multi-head attention block is illustrated in Fig.~\ref{fig5}. Similar to GRU blocks, half of the attention heads are configured as static, while the other half are dynamic. 
Similar dynamic linear blocks are employed not only for the output projections but also for the query, key, and value projections before multi-head partitioning.

Other network components and parameters follow the same configuration as in the FTF-Net. Specifically, all convolutional layers use a kernel size of (2, 3) and a stride of (1, 2) along time and frequency, respectively. 
In grouped GRUs, the dimensionality of the hidden state is set equal to the number of input channels. In the time transformer, trapezoidal masking is utilised to ensure causality and to limit the effective context length to a maximum of 1 second.

It is worth noting that while the proposed dynamic framework is validated on FTF-net, its applicability is broader than this architecture.
The dynamic modules are designed for the most widely used network components and can be applied, individually or in combination, to any architectures that incorporate these blocks.
The contributions of the above modules to multiply-accumulate operations per second (MACs/s) reduction, along with their corresponding reduction percentages, are presented in Table~\ref{tab:table1}, in the case that dynamic components are slimmed for all frames (ratio of frames activating the dynamic components is 0).

\begin{table}[th]
\centering
\begin{threeparttable}
\caption{Contribution of Modules to MAC Reduction ($\Delta$ MAC/s)}
\vspace{-0.2cm}
\label{tab:table1}
\centering
\setlength{\tabcolsep}{0.5mm}
{
\begin{tabular}{ |l|l|c|l|c|} \hline
&{\textbf{Module}} & $\Delta$ MAC \& (\%) & \textbf{Module} & $\Delta$ MAC \& (\%)\\ 
  \hline
Conv layers&{Conv$\_$3} & 11.48 M (50\%)& {Deconv$\_$3} & 11.43 M (50\%) \\ 
  \hline
T-Transformer& MHA &  28.84 M (63\%) & GRUs & 9.02 M (44\%)	\\ \hline
F-Transformer  & MHA & 25.17 M (65\%) & GRUs & 24.47 M (59\%) 	  \\ \hline
	\end{tabular}}
\end{threeparttable}
 \vspace{-0.37cm}
\end{table}

\vspace{-0.15cm}
\subsection{Gating Regularisation}
\vspace{-0.15cm}
For the reconstruction term of the training objective, we adopt the multi-resolution STFT loss from \cite{zhao2025studylightweighttransformerarchitectures}, denoted as $L_\mathrm{multi\_res}$. 
Although the Gumbel-Softmax is employed to promote sparsity in the gating vector generation, it alone is insufficient to prevent the model from activating all candidate network components in pursuit of optimal reconstruction performance. This behaviour contradicts the intended objective of dynamic computational cost reduction. One effective approach in prior work involves setting an average target activation ratio for the gating vectors to constrain the overall utilisation of candidate network resources \cite{Veit_2018_ECCV}.
Based on this objective, we propose a standard gating regularisation loss as follows:
\vspace{-0.2cm}
\begin{equation}
\vspace{-0.2cm}
\mathcal{L}_{\text{gate}} = \max\Bigl(0, \textstyle \frac{1}{T} \sum_{t=1}^{T} g_t - \theta \Bigr)^2
\end{equation}
where $t$ indexes the time frame, and $g_t$ denotes the frame-wise gating value that controls the activations of all dynamic blocks. $\theta$ specifies the desired activation ratio for the dynamic network components.
\begin{figure*}[htbp]
    \centering
    \begin{subfigure}[b]{0.33\linewidth}
        \centering
        \includegraphics[width=\linewidth]{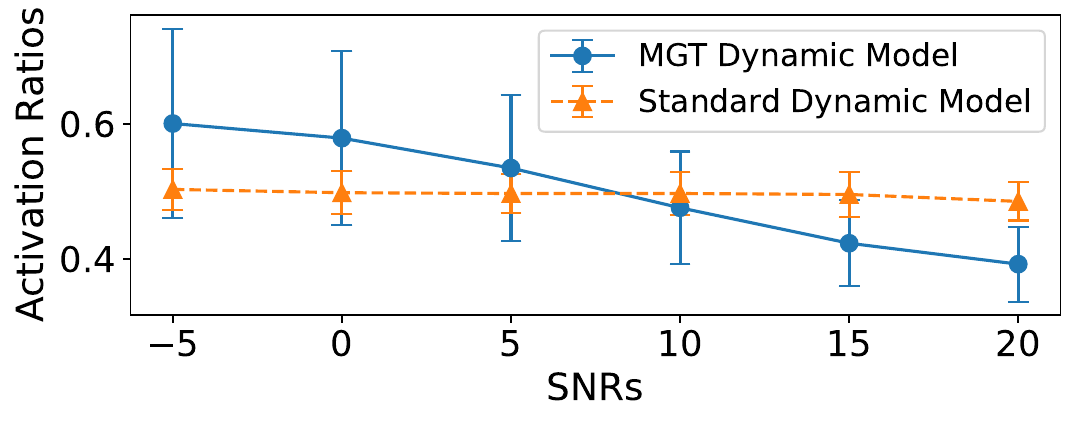}
        \vspace{-0.6cm}
        \caption{Both models across SNRs}
        \label{fig:subfig1}
    \end{subfigure}
    \hfill
    \begin{subfigure}[b]{0.33\linewidth}
        \centering
        \includegraphics[width=\linewidth]{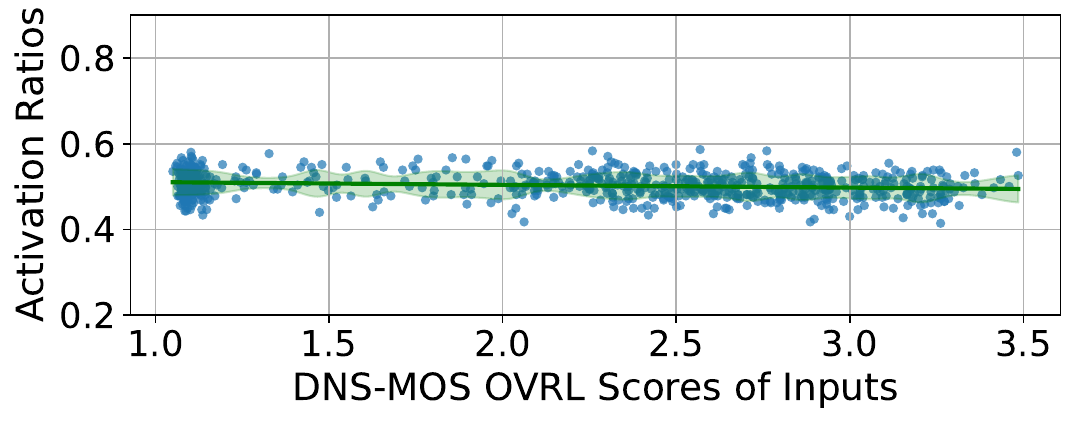}
        \vspace{-0.6cm}
        \caption{Standard dynamic model across OVRL}
        \label{fig:subfig2}
    \end{subfigure}
    \hfill
    \begin{subfigure}[b]{0.33\linewidth}
        \centering
        \includegraphics[width=\linewidth]{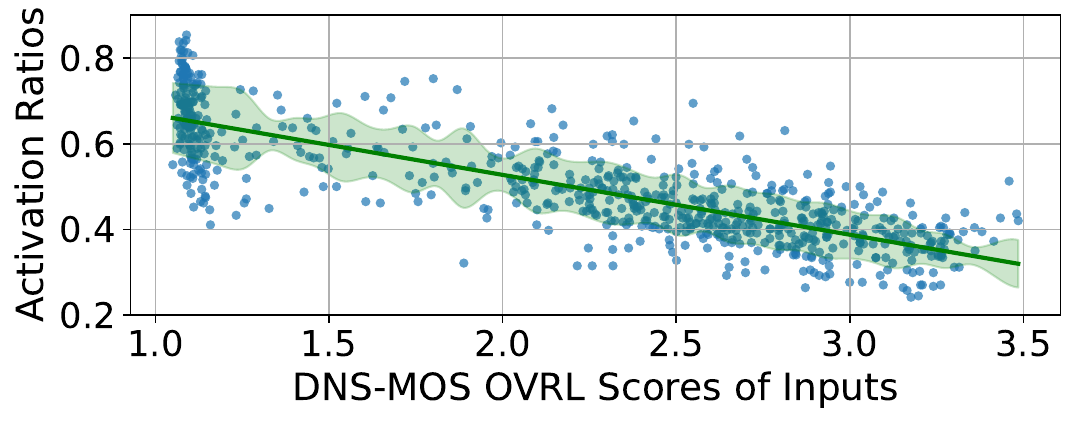}
        \vspace{-0.6cm}
        \caption{ MGT dynamic model across OVRL}
        \label{fig:subfig3}
    \end{subfigure}
    \vspace{-0.65cm}
    \caption{Activation ratios of two proposed dynamic models across SNRs and OVRL on the test dataset. In (a), dots and triangles indicate the mean ratios for each SNR, while vertical error bars denote the standard deviations. Scatter plots displaying ratios of samples along with fitted lines and smoothed error bands, illustrating the trend and confidence intervals in (b) and (c).}
    \label{fig7}
    \vspace{-0.5cm}
\end{figure*}

Although this regularisation term encourages the model to behave dynamically, it relies solely on the trade-off between reconstruction and the utilisation of dynamic components.
without providing explicit guidance regarding input qualities. To address this, we extend the regularisation with a metric-guided formulation: 
\vspace{-0.2cm}
\begin{equation}
\vspace{-0.2cm}
\mathcal{L}_{\text{gate\_MGT}} = \max\Bigl(0, \textstyle \frac{1}{T} \sum_{t=1}^{T} g_t - \theta_m \Bigr)^2
\end{equation}
where $\theta_m$ denotes the target average activation ratio, which is input-signal-dependent and obtained by linearly mapping the input OVRL score of DNS-MOS (P.835) \cite{reddy2022dnsmos}. The mapping is defined as:
$\theta_m = \lambda(5-m)/4$,
where $(5-m)/4$ represents the normalisation of the OVRL score $m$. $\lambda$ is a scaling factor,
which can be configured to tailor the $\theta_m$, based on various practical requirements.
This conditioning guides the model to learn to allocate computational resources appropriately—assigning higher activation ratios for severely degraded inputs and lower ratios for cleaner ones. 
Notably, since the metric scores are only used in guided training, no additional computational overhead is introduced during inference. Here, DNS-MOS OVRL is adopted merely as a proof-of-concept to validate the proposed dynamic framework; alternative metrics like SI-SDR as guidance can be explored in future work.

\vspace{-0.17cm}
\section{Experiments}
\vspace{-0.18cm}
\subsection{Experiment Setup}
\vspace{-0.15cm}
The experiments were conducted using the widely used DNS3 Challenge dataset (DNS3) \cite{reddy21_interspeech}. The training set consists of approximately 140 hours of data synthesised from the wideband English clean speech and noise corpora provided by DNS3, covering a broad range of Signal-to-Noise Ratios (SNRs) from -5 dB to 20 dB in 5 dB increments. The 1-hour unseen test dataset is generated using the same SNR configuration. To ensure causality, no look-ahead is used in experiments.
The STFT is computed using a 512-point window with 50\% overlap, resulting in an algorithm latency of 32 ms. The model is trained using the AdamW optimiser, with a learning rate of $5 \times 10^{-4}$ and a batch size of 8. The exponential decay rates of the optimiser are set to (0.9, 0.99).
\vspace{-0.25cm}
\subsection{Experimental Results}
\vspace{-0.15cm}

Fig.~\ref{fig6} presents the evaluation results of the proposed standard dynamic model, regularised by $\mathcal{L}_{\text{gate}}$, and metric-guided training dynamic model, regularised by $\mathcal{L}_{\text{gate\_MGT}}$, in Extended Short-Term Objective Intelligibility (ESTOI) \cite{taal2010short}, Scale-Invariant Signal-to-Distortion Ratio (SI-SDR) \cite{le2019sdr}, wide-band Perceptual Evaluation of Speech Quality (PESQ) \cite{itut07}, along with the OVRL, SIG, and BAK of DNS-MOS P.835 metrics \cite{reddy2022dnsmos}. The performance curves of models are presented across a range of SNRs from -5 to 20 dB.

The results indicate that the standard dynamic model achieves performance improvements in ESTOI, SI-SDR, and PESQ compared to the zero-activation model by dynamically using more network components. However, it fails to demonstrate similar gains in DNS-MOS metrics, highlighting its limitations in leveraging network resources dynamically.
In contrast, the proposed MGT dynamic model, operating with the same activation ratio and computational complexity, further enhances performance beyond the standard dynamic model and achieves consistently better average metric scores than the zero-activation baseline across all evaluation metrics, especially at low SNRs.
Specifically, at -5 dB SNR, the MGT dynamic model yields a 0.07 increase in DNS-MOS OVRL and a 1.4\% improvement in ESTOI over the standard dynamic model. These results underscore its effectiveness in assessing the quality of input speech and dynamically allocating appropriate network resources accordingly. Interestingly, improvements in the DNS‑MOS BAK are more pronounced than those in the SIG, suggesting that the model’s overall performance gains may be driven by enhanced background noise suppression.
Further, the dynamic models yield comparable performance to the static FTF-Net while requiring only 73\% of the MACs/s. Notably, the MGT model even achieves a slight improvement over FTF-Net in DNS-MOS metrics. This improvement may be related to the training guided on DNS-MOS OVRL scores.

To further demonstrate the effectiveness of metric-guided training and explicitly showcase its capability to assess speech quality, we evaluate the actual activation ratios of the two proposed dynamic models with respect to two different speech quality indicators: SNRs and DNS-MOS OVRL.
As illustrated in Fig.~\ref{fig:subfig1}, the activation ratios of the standard dynamic model remain stable around 50\% across varying SNRs, with no significant variation observed. In contrast, the MGT dynamic model exhibits a marked increase in activation ratios as SNR decreases. In conjunction with the results in Fig.~\ref{fig6}, these higher activation ratios at low SNRs contribute to the MGT dynamic model’s improved performance, while the MGT dynamic model achieves comparable performance at high SNRs, with approximately 10\% fewer activation ratios. % using approximately 10\% fewer network resources on average to 
Furthermore, the larger variations in activation ratios for the MGT model suggest a wider dynamic range and more pronounced differentiation across individual samples, indicating a more discerning dynamic mechanism.
These findings explicitly indicate that the MGT dynamic model's performance gain stems from its correct assessment of speech quality. 
Additionally, scatter plots of activation ratios versus DNS-MOS OVRL scores for both models are presented in Fig.~\ref{fig:subfig2} and Fig.~\ref{fig:subfig3}. For severely distorted samples with low OVRL scores, the MGT dynamic model activates a significantly larger proportion of dynamic network components—up to 90\%. Conversely, for mildly distorted samples with high OVRL scores, activation ratios are relatively low (30-40\%), mirroring the trends observed with SNRs. 
This demonstrates that metric-guided training effectively responds to both intrinsic noise characteristics reflected by SNR and DNS-MOS OVRL.

\begin{table}[th]
\centering
\begin{threeparttable}
\caption{Evaluation Results on Voicebank+Demand Test Dataset}
\vspace{-0.18cm}
\label{tab:table2}
\centering
\setlength{\tabcolsep}{0.0mm}{
\begin{tabular}{ |l|c|c|c|c|c|c|c|c|} \hline
\textbf{Model} & \textbf{Param} & \textbf{MAC} & \textbf{PESQ}  & \textbf{CSIG} &\textbf{CBAK}  & \textbf{COVL} & \textbf{STOI}  & \textbf{SI-SDR}\\ 
  \hline
{Noisy} & - & - & 1.97 & 3.34 & 2.44 & 2.63 & 0.92 & 8.4 	  \\ \hline
FTF-Net \cite{zhao2025studylightweighttransformerarchitectures}      & \textbf{0.14M} & 0.30G & 2.99 & \textbf{4.33} & \textbf{3.61} & \textbf{3.71} & \textbf{0.95}  & 18.8	  \\
{CCFNet+\cite{dang2023first} }                   & 0.62M & 1.47G & \textbf{3.03} & 4.27 & 3.55 & 3.61 & \textbf{0.95} & \textbf{19.1}	  \\ 
MGT-DSN & \textbf{0.14M} & \textbf{0.22G} & 2.98 & 4.31 & \textbf{3.61} & 3.70 & 0.94  & 18.8 \\ 
  \hline
	\end{tabular}}
 \begin{tablenotes}
\item[*] All models are implemented causally, without look-ahead. Best scores within these models are highlighted in bold.

\end{tablenotes}
\end{threeparttable}
 \vspace{-0.2cm}
\end{table} 

To compare the proposed DSN model with SOTA lightweight baselines on a public test dataset, we further train and evaluate it using PESQ, STOI, SI-SDR, and Composite instrumental metrics (CSIG, CBAK, COVL) \cite{hu2007evaluation}, based on the public Voicebank+Demand training and test datasets \cite{valentini2016investigating}. As shown in Table~\ref{tab:table2}, the proposed MGT-DSN model (with 50\% average activation ratio) performs on par with the causal FTF-Net and CCFNet+ baselines \cite{zhao2025studylightweighttransformerarchitectures, dang2023first}, while requiring only 73\% and 15\% of their computational load in MACs/s, respectively. 
For a perceptual appreciation of the MGT-DSN model and additional support to the above discussions, audio samples and figures, with more details about the same results, are available at \href{https://aspire.ugent.be/demos/ICASSP2026HZ/}{https://aspire.ugent.be/demos/ICASSP2026HZ/}. 

\vspace{-0.2cm}
\section{Conclusions}
\vspace{-0.2cm}

We proposed a dynamic network which can provide a further significant reduction in computational complexity by dynamically activating slimmable blocks. 
While validated here on a SOTA lightweight FTF-Net architecture, the dynamic blocks were developed for
the widely used network components, highlighting its applicability to all architectures that employ such components.
With the proposed MGT, the DSN model achieves appropriate network resource allocation guided by input speech quality, indicating its potential to enable guided training in other models.
Further exploration of static versus dynamic component configurations, as well as other metrics for the policy module, is a future direction of research.

\bibliographystyle{IEEEbib}
\bibliography{strings,refs}
% {\setlength{\itemsep}{0pt} % 条目间距
% \bibliographystyle{IEEEbib}
% \bibliography{strings,refs}}

\end{document}